\documentclass[fleqn,twoside]{article}
\usepackage{espcrc2}

\usepackage{graphicx}
\usepackage[figuresright]{rotating}

\newcommand{\AmS}{{\protect\the\textfont2
  A\kern-.1667em\lower.5ex\hbox{M}\kern-.125emS}}

\RequirePackage{xspace}

\usepackage{relsize}
\def\babar{\mbox{\slshape B\kern-0.1em{\smaller A}\kern-0.1em
    B\kern-0.1em{\smaller A\kern-0.2em R}}}
\def\epem       {\ensuremath{e^+e^-}\xspace}

\def\c     {\ensuremath{c}\xspace}

\def\b     {\ensuremath{b}\xspace}

\def\piz   {\ensuremath{\pi^0}\xspace}
\def\pip   {\ensuremath{\pi^+}\xspace}
\def\pim   {\ensuremath{\pi^-}\xspace}
\def\Kp    {\ensuremath{K^+}\xspace}
\def\Km    {\ensuremath{K^-}\xspace}
\def\KS    {\ensuremath{K^0_{\scriptscriptstyle S}}\xspace}

\def\D       {\ensuremath{D}\xspace}
\def\Dbar    {\kern 0.2em\overline{\kern -0.2em D}{}\xspace}
\def\Db      {\ensuremath{\Dbar}\xspace}
\def\Dz      {\ensuremath{D^0}\xspace}
\def\Dzb     {\ensuremath{\Dbar^0}\xspace}
\def\Dstarz  {\ensuremath{D^{*0}}\xspace}

\def\B       {\ensuremath{B}\xspace}
\def\Bbar    {\kern 0.18em\overline{\kern -0.18em B}{}\xspace}

\def\BB      {\ensuremath{B\Bbar}\xspace} 
\def\Bz      {\ensuremath{B^0}\xspace}
\def\Bzb     {\ensuremath{\Bbar^0}\xspace}
\def\BzBzb   {\ensuremath{\Bz {\kern -0.16em \Bzb}}\xspace}
\def\Bu      {\ensuremath{B^+}\xspace}
\def\Bub     {\ensuremath{B^-}\xspace}

\def\Bp      {\ensuremath{\Bu}\xspace}
\def\Bm      {\ensuremath{\Bub}\xspace}

\def\BpBm    {\ensuremath{\Bu {\kern -0.16em \Bub}}\xspace}
\def\Bs      {\ensuremath{B_s}\xspace}

\def\beq	{\begin{equation}}
\def\eeq	{\end{equation}}

\title{CP violation and CKM matrix elements at the B factories}

\author{J. Chauveau\address[LPNHE]{
Laboratoire de Physique Nucl\'eaire et de Hautes Energies, \\
IN2P3/CNRS, Universit\'e Pierre et Marie Curie-Paris6} 
\thanks{On behalf of the \babar\ Collaboration} }

\begin{document}

\begin{abstract}
Recent measurements at the B factories of CKM matrix elements affecting 
CP-violation are reviewed. The emphasis is on the unitarity triangle.
Some aspects of the charm and $\tau$ sectors are mentioned. 
\vspace{1pc}
\end{abstract}

\maketitle

\section{INTRODUCTION}
In the standard model (SM) the Cabibbo-Kobayashi-Maskawa (CKM) matrix
element $V_{ij}$ is the quark mixing coupling factor in the weak 
charged current connecting the i-th u-type quark to the j-th d-type quark 
and the $W$~\cite{Cabibbo,KM}.
With three generations four real parameters are needed to describe 
the unitarity CKM matrix $V$, among which there is an irreducible phase which 
governs all CP violating phenomena. The unitarity relation built from the $d$ 
and $b$ columns of $V$ defines a triangle in the complex plane. It is 
convenient to normalize the sides and to measure the phases with respect to 
$V_{cd}V^*_{cb}$, obtaining the unitarity triangle called the UT in the 
following. Its apex represents the complex number:
\beq
z = \overline{\rho}+ i \overline{\eta} \equiv -\frac{V_{ud}V_{ub}^*}{V_{cd}V_{cb}^*}
\eeq
whose imaginary part $\overline{\eta}$ governs CP violation. In other words, 
CP non-conservation means that the triangle is not squashed. Its angles
$\alpha$ ($\phi_2$) at the apex, $\gamma$ ($\phi_3$) at the origin and $\beta$ ($\phi_1$)
differ from 0 or 180$^{\circ}$. Use of the Wolfenstein 
parameterization~\cite{Wolfenstein} is customary with the definitions:
\begin{eqnarray}
\lambda^2 \equiv \frac{|V_{us}|^2}{|V_{ud}|^2+|V_{us}|^2} \\
A^2\lambda^4 \equiv \frac{|V_{cb}|^2}{|V_{ud}|^2+|V_{us}|^2}. 
\end{eqnarray}
$\lambda$ is the sine of the Cabibbo angle and $A$ is a real number 
of order unity. $\gamma = arg(z)$ is the opposite of the phase of $V_{ub}$. \par

Many measurements contrain the apex of the UT. Dedicated teams produce global fits. 
I use the results of the CKMfitter group~\cite{CKMfitter,SDG}. 
They are in fair agreement with those of the UTfit
group~\cite{utfit}, presented in another lecture~\cite{Ciuchini} at this Conference.

\begin{figure*}[htb!]
\hspace*{10mm}\includegraphics[width=125mm]{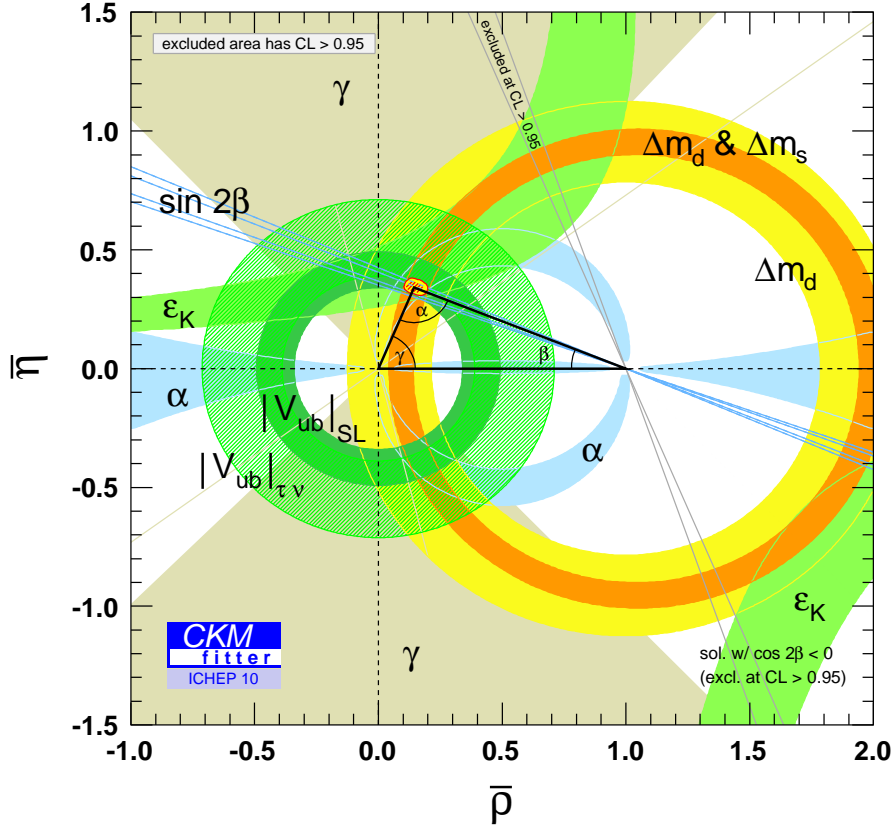} 
\caption{\label{fig:CKMfitter} Figure from reference~\cite{CKMfitter}. 
The constraints in the ($\overline{\rho}, \overline{\eta}$) plane and the global CKM fit
the 68\% confidence level of which is shown as the hashed region.
The $|V_{ub}|$ constraint is split into two contributions: that from $\B$ semileptonic decays (dark), and that from 
$\B \to \tau \nu$ (hashed). 
}
\end{figure*}

The global fit pictured in Figure~\ref{fig:CKMfitter} and summarized in 
Table~\ref{tab:CKMfitter} shows the agreement of the experiments 
with the SM.  While the Cabibbo angle is precisely measured, the accuracy on 
the other parameters ranges between 4 and 20\%.
In the following I review a subset of recent measurements 
of the angles and the sides of the UT.

\begin{table}[htb]
\caption{Global fit results from CKMfitter~\cite{CKMfitter,SDG}. 
The uncertainties are quoted at the 68\% confidence level.}
\label{tab:CKMfitter}
\newcommand{\m}{\hphantom{$-$}}
\newcommand{\cc}[1]{\multicolumn{1}{c}{#1}}
\renewcommand{\tabcolsep}{2pc} 
\renewcommand{\arraystretch}{1.2} 
\begin{tabular}{@{}ll}
\hline
\hline
$A$              & $0.8184^{+0.0094}_{-0.0311}$ \\
$\lambda$        & $0.22512^{+0.00075}_{-0.00075}$ \\
$\overline{\rho}$& $0.139^{+0.027}_{-0.023}$ \\
$\overline{\eta}$& $0.342^{+0.016}_{-0.015}$ \\
\hline
\end{tabular}
\end{table}

\section{THE ANGLES}
The angles $\beta / \phi_1$ and $\alpha / \phi_2$ have been well measured
since 2007. In contrast, progress is continously made in the determination of the angle $\gamma/ \phi_3$.
\subsection{Recent results on the angle $\gamma / \phi_3$}
\begin{figure}[htb]
\includegraphics[width=37.5mm]{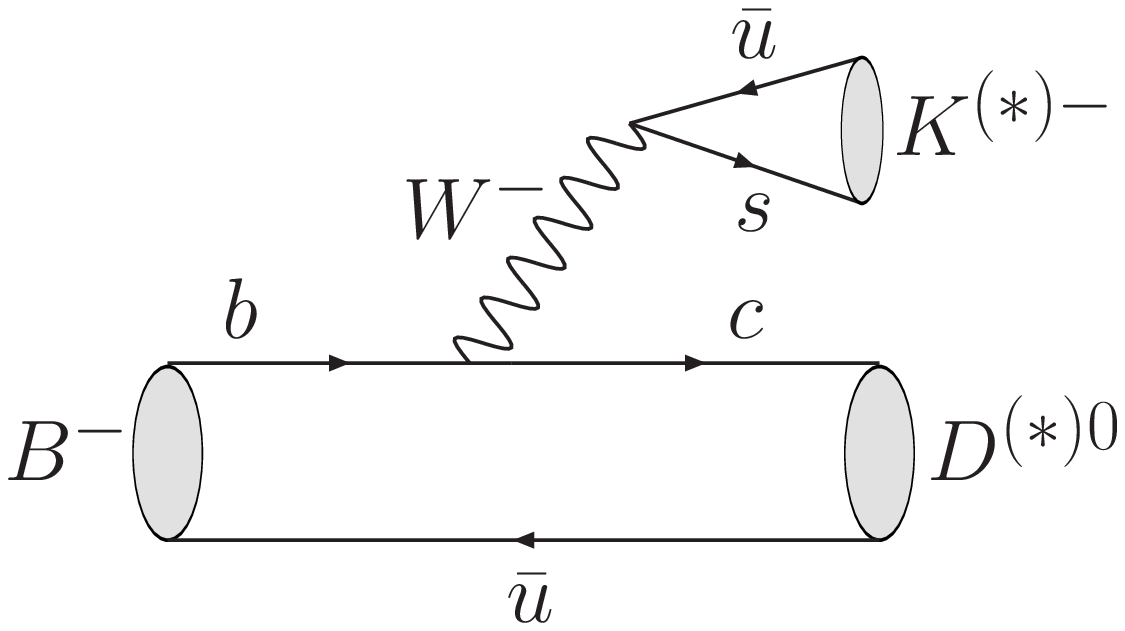}\includegraphics[width=37.5mm]{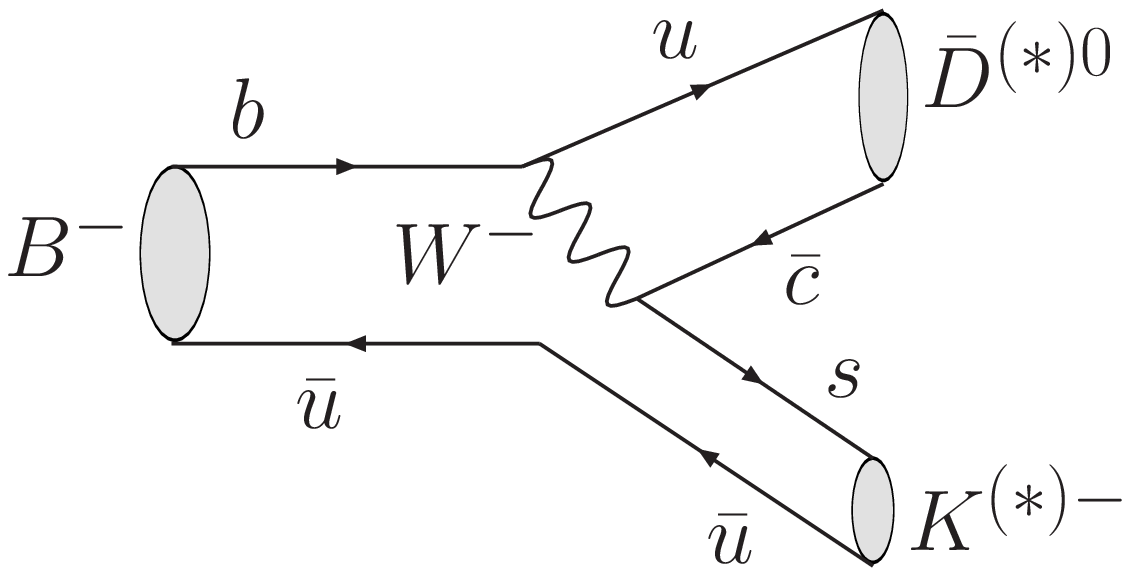}
\caption{ADS method. Feynman diagrams of the $b \to u$ and $b \to c$ processes which interfere in the decay 
$\Bm \to \underline{D}\Km$.}
\label{fig:Feyn}
\end{figure}

Three methods~\cite{GLW,ADS,GGSZ} are exploited to measure the angle 
$\gamma / \phi_3$ in charged (or neutral) $\B \to \D^{(*)}K^{(*)}$ decays mediated by tree
amplitudes. The relative phase of $V_{ub}$ and $V_{cb}$ appears in the
interference of the diagrams of Figure~\ref{fig:Feyn}, when they lead 
to the same final state $\underline{D} K$\footnote{We generically use $\underline{D}$ for $\D$/$\Db$ as both feed the final~state.}. 
In the GLW method~\cite{GLW}, the $\Dz$ CP-eigenstates are selected. 
In the ADS method~\cite{ADS},  the same final state is reached when 
the $b \to c$ transition is followed by a Doubly Cabibbo Suppressed $\D$ decay (DCSD) e.g. $\Dz \to \Kp\pim$ or 
the $b \to u$ transition is followed by a Cabibbo allowed $\D$ decay (CA) e.g. $\Dzb \to \Kp\pim$. 
In the GGSZ method~\cite{GGSZ} the rich structure (Dalitz plot) of a flavor-blind 3-body $\D$ decay is the key. 
Large uncertainties affect the measurements because the rates are low or the two decay paths have disparate magnitudes. 
As the CA and DCSD $\D$ decay rates are precisely known, the measured interference patterns are governed by 
$r_B$ the ratio of the $b \to u$ and $b \to c$ magnitudes, $\delta_B$ and $\delta_D$ the strong phase shifts
in the $\B$ and $\D$ decays of the 2 paths, and $\gamma / \phi_3$. Recent results have been obtained using 
ADS and GGSZ by the \babar\ and Belle collaborations, which we now summarize. \par

\begin{figure*}[htb!]
\begin{tabular}{ccc}
\includegraphics[width=42mm]{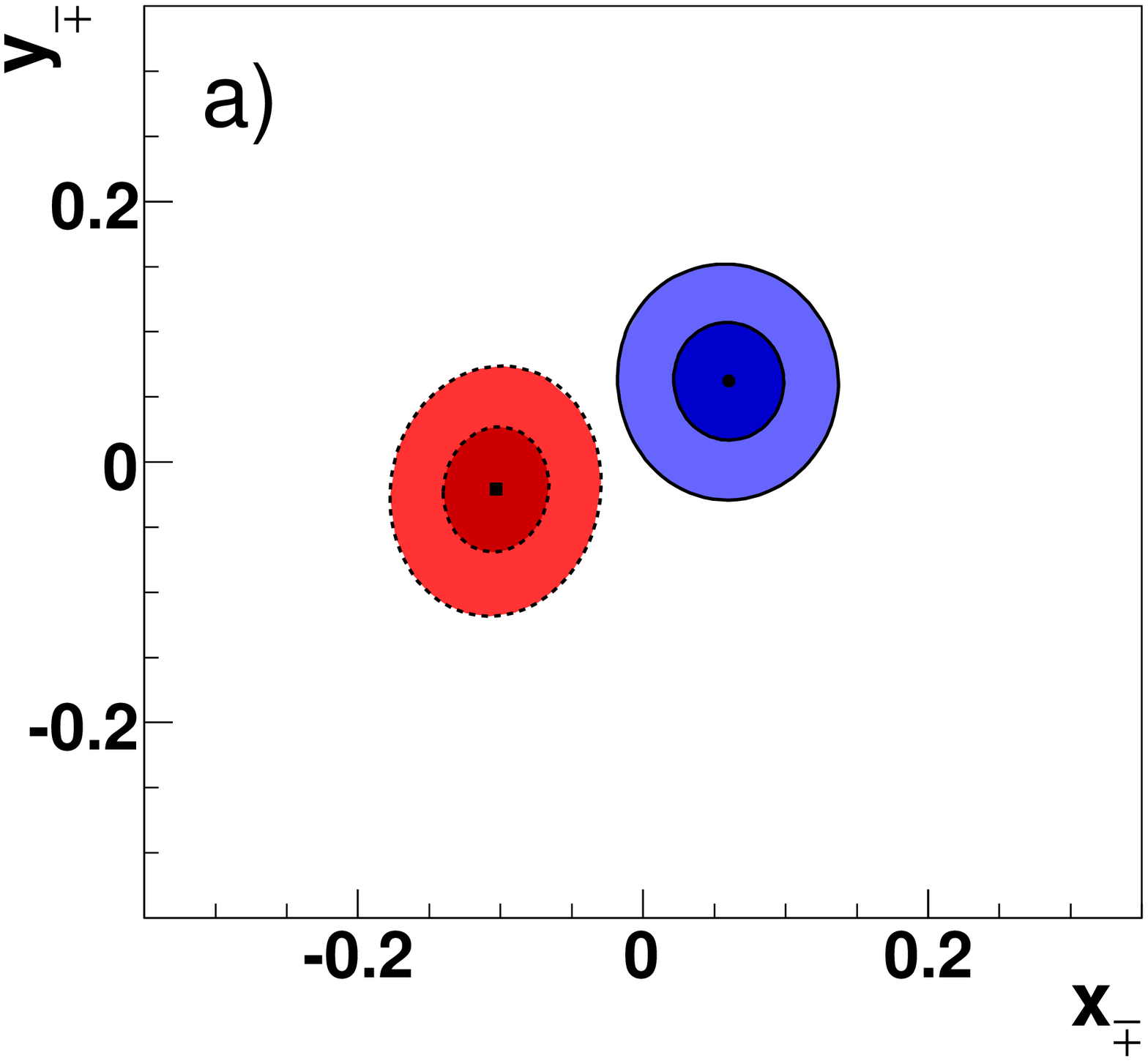}&
\includegraphics[width=42mm]{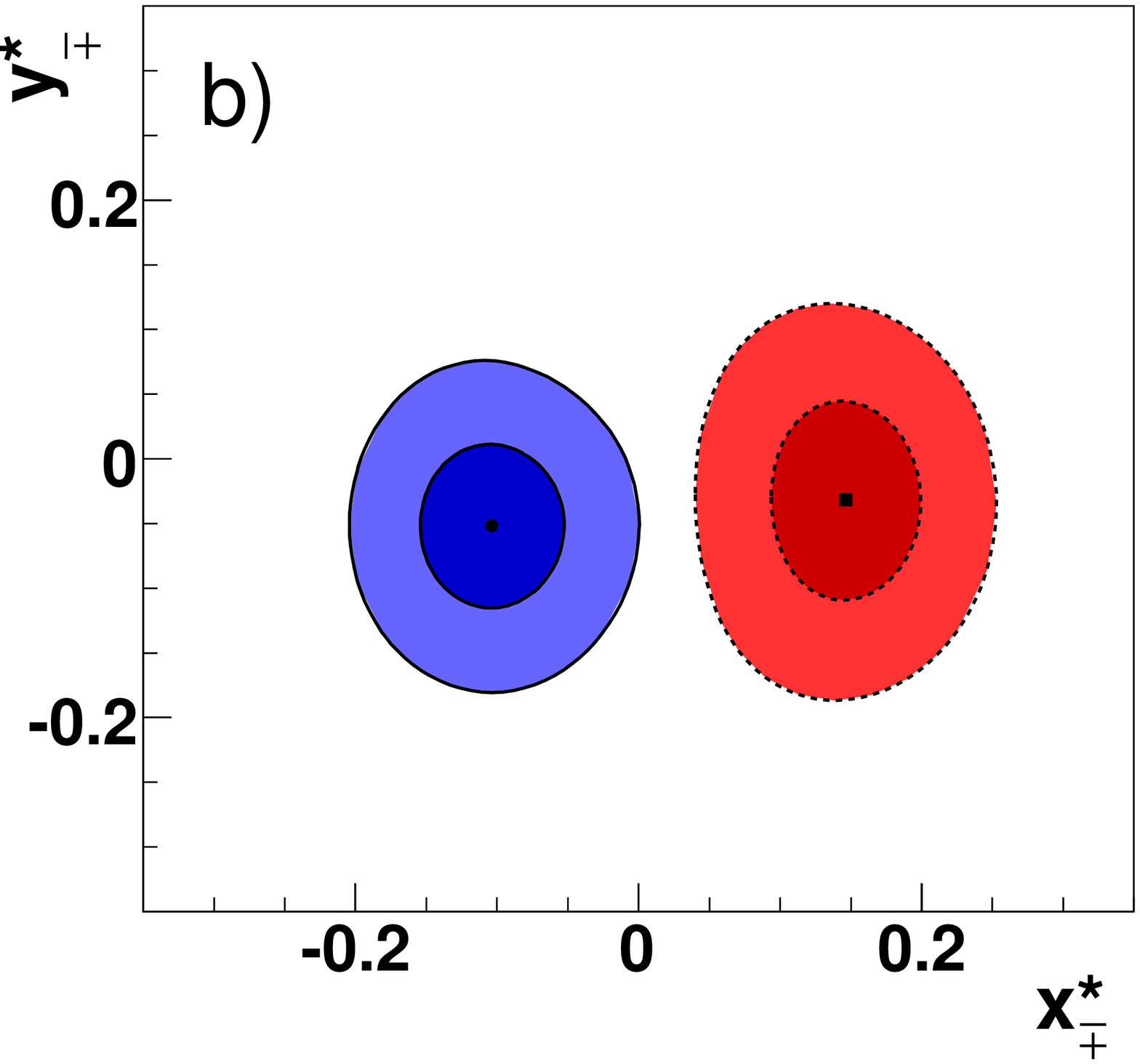} &
\includegraphics[width=42mm]{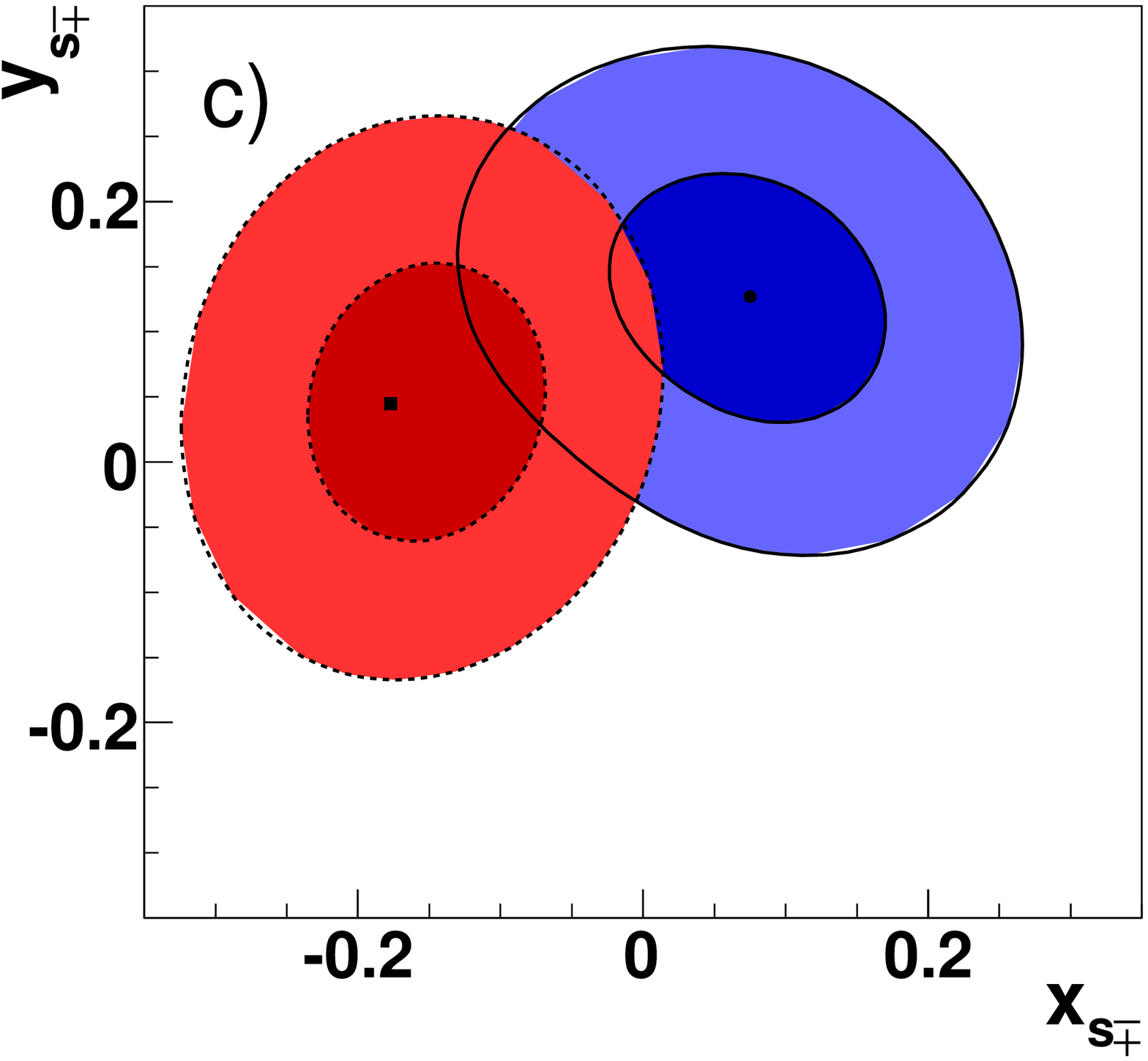} \\
\end{tabular}
\caption{\label{fig:babarGGSZ} Figure from reference~\cite{Fernando}. 
Contours at $39.3\%$ (dark) and $86.5\%$ (light) 2-dimensional confidence levels in the cartesian
coordinate planes for (a) $\D K$, (b) $D^*K$, and (c) $DK^*$, corresponding to one- and two-standard deviation regions 
(statistical only), for $\Bm$ (solid lines) and $\Bp$ (dotted lines) decays. 
The two contours in each picture would coincide if CP were conserved. 
The angle between the (to-be-imagined) straight lines joining the $(x=0, y=0)$ point to the 
two contour centers is equal to $2 \gamma$.  
}
\end{figure*}

The observables in the ADS method are the CP-averaged  $b \to u$ / $b \to c$ ratio of decay rates $R_{ADS}$, and
the CP asymmetry $A_{ADS}$. The sensitivity to the angle $\gamma / \phi_3$ is apparent from the relations:
\small
\begin{eqnarray}
R_{ADS}&=&r_B^2+r_D^2+2 r_B r_D \cos(\delta_B + \delta_D) \cos \gamma \\
A_{ADS}&=&[2 r_B r_D \sin(\delta_B + \delta_D) \sin \gamma]/R_{ADS}.
\end{eqnarray}
\normalsize
\babar~\cite{Lees} has recently measured these quantities in the three channels 
$\Bm \to K^- + \underline{\Dz}, \Dstarz [\to (\piz$ or $\gamma) +\underline{\Dz}]$ (Table~\ref{tab:ADS}) and 
observed indications of signals at the two standard deviation significance level in the first two.
\begin{table*}[htb]
\caption{ADS results from \babar~\cite{Lees}.}
\label{tab:ADS}
\begin{tabular}{@{}llll}
\hline
channel               & $ R_{ADS}\times 10^2 $ & significance & $A_{ADS}$  \\
\hline
$DK$		      & $1.1 \pm 0.5 \pm 0.2 $ & $2.1 \sigma$ & $-0.86 \pm 0.47 ^{+0.12}_{-0.16}$	\\
$\Dstarz K, \Dstarz \to \Dz\piz$  & $1.9 \pm 0.9 \pm 0.4 $ & $2.2 \sigma$ & $+0.77 \pm 0.35 \pm 0.12$ 		\\
$\Dstarz K, \Dstarz \to \Dz\gamma$& $1.3 \pm 1.4 \pm 0.8 $ & $-$          & $+0.36 \pm 0.94 ^{+0.25}_{-0.41}$	\\
\hline
\end{tabular}
\end{table*}
The GGSZ method~\cite{GGSZ} has been used to interpret the measurements of the same $\B$ decays and $\B \to \Dz K^*(892)$ 
where the $\Dz$ meson decays 
into 3-body channels $\underline{\Dz} \to \KS \pip\pim $~\cite{belleGGSZ} and 
$\underline{\Dz} \to \KS \pip\pim,\  \KS\Kp\Km$~\cite{Fernando} using respectively 657 and 468 millions $\BB$ pairs. 
The resulting {\it cartesian coordinates}
\beq
x^{\pm} \equiv r_B \cos(\delta_B \pm \gamma);\ y^{\pm} \equiv r_B \sin(\delta_B \pm \gamma) 
\eeq
provide the best determination to date of the angle $\gamma / \phi_3$. The \babar\ measurements are shown on 
Figure~\ref{fig:babarGGSZ}. Table~\ref{tab:GGSZ} lists the measurements of $\gamma / \phi_3$ 
and the ancillary parameters from \babar\ and 
Belle. The results of the two experiments are compatible, both claim evidence for direct CP violation ($\gamma \not= 0$)
at the 3.5 standard deviation level.  The combination by the CKMfitter group of all measurements gives 
$\gamma_{comb}=70^{+14\ \circ}_{-21}$  to be compared to the global CKM fit (leaving the $\gamma$ measurements aside) result
$\gamma_{CKMfit}=67.4\pm 3.9 ^{\circ}$.

\begin{table*}[htb]
\caption{GGSZ results from \babar~\cite{Fernando} and Belle~\cite{belleGGSZ}.}
\label{tab:GGSZ}
\begin{tabular}{@{}lllll}
\hline
Parameter                  & Experiment & $68\%$ CL (stat.) & exp. systematics & Dalitz model syst.       \\
\hline
$\gamma / \phi_3 (^{\circ})$ & \babar	&  $68^{+15}_{-14}$         & $\pm 4$      & $\pm 3$	          \\
                           & Belle      &  $78.4^{+10.8}_{-11.6}$   & $\pm 3.6$    & $\pm 8.9$            \\
$r_B (DK)$                 & \babar	&  $0.096\pm2.9$            & $\pm 0.005$  & $\pm 0.004$          \\
			   & Belle      &  $0.160^{+0.040}_{-0.038}$& $\pm 0.011$  & $^{+0.050}_{-0.010}$ \\
$r_B (D^*K)$		   & \babar	&  $0.133^{+0.042}_{-0.039}$& $\pm 0.013$  & $\pm 0.003$          \\
			   & Belle      &  $0.196^{+0.072}_{-0.069}$& $\pm 0.012$  & $^{+0.062}_{-0.012}$ \\
$r_B^{eff}(DK^*)$	   & \babar	&  $0.149^{+0.066}_{-0.062}$& $\pm 0.026$  & $\pm 0.006$	  \\
$\delta_B (DK)(^{\circ})$  & \babar	&  $119^{+19}_{-20}$	    & $\pm 3$	   & $\pm 3$		  \\ 
			   & Belle      &  $136.7^{+13.0}_{-15.8}$  & $\pm 4$      & $\pm 22.9$           \\
$\delta_B (D^*K)(^{\circ})$ & \babar	&  $-82 \pm 21$             & $\pm 5$      & $\pm 3$              \\
			   & Belle      &  $-18.1^{+18.0}_{-18.6}$  & $\pm 3$	   & $\pm 22.9$           \\
$\delta_B^{eff}(DK^*) (^{\circ})$& \babar & $111 \pm 32$		    & $\pm 11$     & $\pm 3$              \\
\hline
\end{tabular}
\end{table*}

\subsection{The $\beta / \phi_1$ and $\alpha / \phi_2$ angles}
Using all measurements of $\B \to$ charmonium + $K_{S,L}$ the heavy flavor averaging group (HFAG) determines~\cite{hfag}
$\sin 2\beta=0.673 \pm 0.023$ which translates into either $\beta / \phi_1 = 21.1\pm 0.9 ^{\circ}$ or $69.9 \pm 0.9 ^{\circ}$.
The former value is slightly preferred experimentally and matches the other contraints in the
global CKM fit wich validates the SM. It has to be noted that the most recent measurements done using Dalitz-plot amplitude analyses
(e.g for $\B \to \pip\pim\KS$~\cite{Alejandro,BelleKKKS}) determine the angle $\beta / \phi_1$ 
without trigonometric ambiguity. However they are less precise. \par
The $\alpha / \phi_2$ angle is precisely determined by the $\B \to \pi \pi$, $\rho\pi$ and $\rho\rho$ decays, 
analyzed using isospin relations to disentangle tree and penguin decay paths. 
Combining all experimental results the CKMfitter group finds 
$\alpha_{comb} = 89^{+4.4\ \circ}_{-4.2} $ which compares well to the global fit (leaving the $\alpha$ 
measurements aside) result, $\alpha_{CKMfit}=97.5^{+1.6\ \circ}_{-8.1}$.\par
\vspace*{5mm}
In summary the measurements of the angles $\beta / \phi_1$ and $\alpha/ \phi_2$ alone constrain the apex of the UT 
enough to establish that CP is violated in $\B$ decays. The determination of the $\gamma/ \phi_3$ angle is not yet
accurate enough to further constrain the KM model. The GLW, ADS and GGSZ methods are effective at LHCb. With them
very clean measurements of the $\gamma / \phi_3$ angle in pure tree processes will be obtained which will establish a solid basis for 
new physics searches in $\gamma$-sensitive decays proceeding via loop diagrams.    

\section{THE SIDES}
I refer to the recent review of the determination of $|V_{ub}|$ and $|V_{cb}|$ from  the 
semileptonic $\B$ meson decays at the B-factories~\cite{BobK} and the compilation of the 
Review of Particle Properties~\cite{bslinpdg}. With increased statistics, well established 
analysis techniques still progress. Tagged samples of $\BB$ events where the {\it other} $\B$ meson is reconstructed 
now contribute. Event selection algorithms have been devised that increase the acceptance of $\b \to u \ell \nu$
decays over wide kinematical ranges. Inclusive and exclusive semileptonic decays are both relevant. The former 
are modeled theoretically using Operator Product Expansions (OPE) by expressions in terms 
of $\alpha_s$ and $\frac{\Lambda_{QCD}}{m_b}$ with useful commonalities with those describing the $\B \to X_s \gamma$ 
inclusive radiative decays.  The exclusive semilptonic decays rely on sets of  
form factors conform to Heavy Quark Symmetry and lattice QCD (LQCD) computations. 
$|V_{ub}|$ is also constrained by the purely leptonic $\B \to \tau \nu$ decay,
simpler to model but with a dependence on the $\B$ meson decay constant to be computed from LQCD. \par
\subsection{$V_{cb}$}
The current exclusive $\B \to D^* \ell \nu$ measurements combined with a recent unquenched lattice calculation~\cite{BernardLQCD}
lead to $|V_{cb}|= (38.7 \pm 0.9 \pm 1.0)\times 10^{-3}$, where the experimental uncertainty is quoted first followed 
by that of the LQCD calculations. A new tagged analysis from \babar~\cite{babartaggedDlnu}
of $\B \to D \ell \nu$ has a significantly improved accuracy. With an unquenched 
lattice form factor calculation~\cite{LQCD-Dlnu}, it leads to $|V_{cb}|= (39.1 \pm 1.4 \pm 1.3)\times 10^{-3}$. 
Global fits to the inclusive decay rates and $\simeq 60$ moments of the lepton energy and hadronic mass 
spectra~\cite{babarprdclnu} as well the $\B \to X_s \gamma$ photon energy spectrum~\cite{bellesgamma} are performed to
extract $|V_{cb}|$, hadronic coefficients and the \b quark mass within a given renormalization scheme. 
The average~\cite{BobK,hfag} for inclusive decays is  $|V_{cb}|= (41.9 \pm 0.4 \pm 0.6)\times 10^{-3}$. The agreement
between the $|V_{ub}|$ determinations from exclusive and inclusive decays is marginal. The uncertainties have to be
scaled up in order to quote the average $<|V_{cb}|>= (40.9 \pm 1.0)\times 10^{-3}$.
\subsection{$V_{ub}$}
Because the rates are suppressed, the background from $\b \to \c$ processes is high and the inclusive measurements are
harder. Strict selections are implemented which reduce the accepted phase space. Therefore the applicability of 
OPE is impaired unless one uses a {\it shape function}~\cite{BobK,bslinpdg}. Weak annihilations contribute. Furthermore, a recent
calculation~\cite{BNLP} shows that QCD effects at the NNLO are sizeable.      
With high statistics however, innovative analysis techniques are implemented. For instance, with multivariate techniques
Belle~\cite{bellevubinclusive} accesses 90\% of the available phase space. Inclusive determinations  
average~\cite{BobK} to: $|V_{ub}|=(4.37 \pm 0.16 \pm 0.20 \pm 0.30)\times 10^{-3}$, 
where the first uncertainty is experimental, 
the second is from the spread of theoretical results and the third is the NNLO effect. \par
A preliminary result from \babar~\cite{babarpilnu} on  $\B\to\pi\ell\nu$ comes out of a fit of the differential 
branching fraction as a function of $q^2$ which uses data points from the experiment as well as from a lattice 
calculation~\cite{MILCpilnu}. 
The result, $|V_{ub}|=(2.95\pm0.31)\times 10^{-3}$ where the uncertainty combines the experimental and theoretical 
contributions, is lower by $\simeq 20\%$ from previous averages. There again and more significantly than for 
$|V_{cb}|$, the exclusive and inclusive averages are far apart. 
Recent findings tend to increase the discrepancy. This should be kept in mind when considering  
that the  global CKM fits inputs are the averages of the exclusive and inclusive determinations of $|V_{ub}|$ and $|V_{cb}|$.\par
A stronger feature involving  $|V_{ub}|$ comes from the determination of the $\B \to \tau \nu$ branching fraction.
A higher value than that from the semileptonic decays is preferred by the data from \babar\ and Belle and this bring
a much discussed tension in the global CKM fits with the constraint from $\sin 2\beta$. I refer to the
thorough description of that issue~\cite{Schwanda} at this workshop.
\subsection{$V_{ts}/V_{td}$}
The $\frac{V_{td}V_{tb}^*}{V_{cd}V_{cb}^*}$ side of the UT remains to be discussed. The most accurate relevant contraints 
come from measurements~\cite{CKMfitter} of $|\frac{V_{ts}}{V_{td}}|$, currently
dominated by $\Bs$ mixing~\cite{bsmixcdf}. A new \babar\ measurement~\cite{babarxsgamma}
of the inclusive decays to $\B \to X_{s,d} \gamma$ brings a contribution with independent systematic 
uncertainties: $|\frac{V_{ts}}{V_{td}}|=0.199 \pm 0.022 \pm 0.024 \pm 0.002$, where the last uncertainty after the 
statistical and the systematic components is from theory.  
  
\section{OTHER TOPICS}
For lack of time in Capri and lack of place here, I do not describe two topics.
Recent attempts to check the unitarity of the first row of the CKM matrix using strange and non-strange 
$\tau$ decays are conclusive with exclusive ($\pi \nu$ and K$\nu$) decays; but not yet with inclusive ones.
To date, there  no evidence of CP violation in $D$ meson mixing. 
 
\section{SUMMARY AND OUTLOOK}
CP violation in $\B$ meson decays has been established. The KM model works to explain CP violation phenomena
observed with quarks. The global CKM fits reveal some tensions. The contrary would be suspicious... None 
are overwhelming. Each should be scrutinized. There is room for New Physics which could appear as corrections of 
order 10\% to flavor parameters. Many measurements are statistically limited. 
Now is the time of the hadron machines (TeVatron, LHCb). Hopefully New Physics will unveil. However \epem colliders 
are invaluable to pursue a comprehensive experimental program. Still the 
antimatter problem remains unexplained.
  
\section{ACKNOWLEDGEMENTS}
I am very grateful to have been invited at the Capri workshop. Many thanks to the organisers, the \babar\ Collaboration 
and to LPNHE and IN2P3/CNRS. Special thanks to the CKMfitter group for sharing their most recent results.

\end{document}